\documentclass[a4paper,11pt]{article}
\usepackage{pos}

\usepackage{cleveref}
\usepackage{xspace}
\usepackage[justification=centering]{subcaption}
\usepackage{booktabs}
\usepackage{multirow}

\makeatletter
\g@addto@macro\bfseries{\boldmath}
\makeatother

\title{Measuring Light-Meson Resonances in the $\omega\pi^-\pi^0$ and $K_S^0 K^-$ Final States at COMPASS}

\author*[a]{Julien Beckers}
\author[a]{Philipp Haas}
\onbehalf{on behalf of the COMPASS Collaboration}

\affiliation[a]{Physics Department, School of Natural Sciences, Technical University of Munich,\\
  James-Franck-Str. 1, Garching bei München, Germany}

\emailAdd{julien.beckers@tum.de}
\emailAdd{philipp.haas@tum.de}

\abstract{
COMPASS is a multi-purpose fixed-target experiment at the CERN SPS. One of its main goals is to probe
the strong interaction at low energies by studying the excitation spectrum of light mesons in diffractive scattering reactions of a $190\ \text{GeV}/c$ $\pi^-$ beam.
The analysis is done by first decomposing the data into partial-wave amplitudes with well-defined quantum numbers, and second, 
extracting meson resonance parameters from these amplitudes.

We have collected the world's largest datasets of various final states. In this talk, we will focus on two of them: $\omega\pi^-\pi^0$ and $K_S^0 K^-$. They allow us to study light isovector mesons with spin, parity, and $C$-parity $J^{PC} = J^{++}$ and $J^{-+}$, i.e.\ $a_J$ and $\pi_J$ mesons. We will discuss the analysis and present new measurements of resonance parameters of several light mesons.

The main focus of the $\omega\pi^-\pi^0$ analysis lies in the investigation of the nature of the $\pi_1(1600)$. Being a good candidate for the lightest hybrid meson, it is expected to predominantly decay into $b_1(1235)\pi$. In addition, the $\omega\pi^-\pi^0$ final state also gives access to other decay modes of the $\pi_1(1600)$, further testing theory predictions, and to a range of other $a_J$ and $\pi_J$ mesons.

In the $K_S^0 K^-$ final state, only $a_J$ mesons with even spin $J$ appear, due to the high beam energy of COMPASS. This allows for an exclusive study of these mesons, up to high invariant masses, verifying the existence of several states claimed by other experiments and measuring their parameters.

}

\FullConference{The 21st International Conference on Hadron Spectroscopy and Structure (HADRON2025)\\
 27 - 31 March, 2025\\
Osaka University, Japan\\}

\newcommand{\Ks}{\ensuremath{K_S^0}\xspace}
\newcommand{\Km}{\ensuremath{K^-}\xspace}
\newcommand{\KsK}{\ensuremath{\Ks \Km}\xspace}

\newcommand{\oPP}{\ensuremath{\omega \pi^0 \pi^-}\xspace}

\newcommand{\mX}{\ensuremath{m_X}\xspace}
\newcommand{\tpr}{\ensuremath{t'}\xspace}

\newcommand{\JP}{\ensuremath{J^{P}}\xspace}
\newcommand{\JPM}{\ensuremath{J^{P}M}\xspace}

\newcommand{\GeVc}{\ensuremath{\text{GeV}/c}\xspace}
\newcommand{\GeVcc}{\ensuremath{\text{GeV}/c^2}\xspace}

\newcommand{\MeVcc}{\ensuremath{\text{MeV}/c^2}\xspace}

\begin{document}
\maketitle

\section{Introduction}\label{sec:intro}


We study the strong interaction by precisely
determining the fundamental properties of hadrons. Exotic hadrons, i.e.\ states beyond the simple $q\bar{q}$ or $qqq$ configurations of the quark model, are particularly interesting.
The COMPASS experiment at CERN has a key role in the search for new, unstable light mesons $X^-$, produced by scattering a $190\ \GeVcc$ pion beam off a liquid-hydrogen target, and decaying into an $n$-body final state,
\begin{equation}\label{eq:process_general}
    \pi^- + p \to X^- + p \to (1 + 2 + ... + n) + p \, .
\end{equation}
Here, we will present a selection of results from analyses of the \KsK and the \oPP final states.

\section{The partial-wave analysis method}\label{sec:methods}

In this Section, we summarize the method of partial-wave analysis, detailed e.g.\ in \cite{compass_PWA_review}.

\subsection{Partial-wave decomposition}\label{sec:methods_pwd}

To study the process in \cref{eq:process_general}, we start by building a model for its intensity differential in the $n$-body phase space $\tau_n$ for fixed invariant mass \mX and reduced four-momentum transfer $\tpr$ \cite{compass_PWA_review} between beam and target. We decompose the total amplitude into amplitudes of partial waves, i.e.\ contributions with specific quantum numbers and decay channels labeled by $a \equiv \JPM \text{[decay]}$,
\footnote{See \cite{compass_PWA_review} for definitions of the spin $J$, parity $P$ and the spin projection $M$. We omit the isospin $I=1$, $C$-parity $C=+1$ and $G$-parity $G=-1$ as well as the reflectivity $\varepsilon=+1$.}
\begin{equation}\label{eq:intensity_decomp}
    \mathcal{I}(m_X,\tpr;\tau_n) = \left| \sum_{a} \, \mathcal{T}_a(m_X,\tpr) \, \psi_a(m_X;\tau_n) \, \right|^2 \ .
\end{equation}
Assuming intermediate resonances $X^-$ dominate the process, we have factorized \cref{eq:intensity_decomp} further into the production and propagation of a resonance $X^-$ and its decay into the final state.
The latter is modeled via decay amplitudes $\psi_{a}(m_X;\tau_n)$ describing the phase-space distribution. In \KsK, these are Wigner $D$-functions. In \oPP, each amplitude is calculated assuming a two-body intermediate state in the decay, the so-called isobar resonance, e.g.\ $X^-\to b_1\pi;\  b_1\to\omega\pi$.
The transition amplitudes $\mathcal{T}_a(m_X,\tpr)$ contain information about the produced states $X^-$. In this first step, we extract their dependence on $m_X$ and $\tpr$ by dividing the data into cells of these variables and performing independent maximum-likelihood fits to the data to estimate the values of $\mathcal{T}_a$. Uncertainties are estimated using the Bootstrap method \cite{bootstrap}.

\subsection{Resonance-model fit}\label{sec:methods_rmf}

In a second step, we identify the resonances $X^-$ contributing to process \cref{eq:process_general} and measure their masses $m_0$ and widths $\Gamma_0$. To this end, we model the $m_X$ dependence of the amplitudes, extracted in the partial-wave decomposition (\cref{sec:methods_pwd}), as a coherent sum of resonant and non-resonant components. Resonances are described by relativistic Breit-Wigner amplitudes,
\begin{equation}\label{eq:BW}
    D_{\mathrm{R}}(\mX;m_0,\Gamma_0) = \frac{m_0 \, \Gamma_0}{m_0^2 - \mX^2 + i \, m_0 \, \Gamma(\mX)} \ ,
\end{equation}
with the dynamic width $\Gamma(\mX)$.
Non-resonant background is modelled phenomenologically \cite{3pi_rmf} via
\begin{equation}
    D_{\text{NR}}(\mX;a,b) = \left( \mX - m_{\text{thr}} \right)^a
    \cdot \, \exp\left[-b \cdot q^2(\mX^2)\right] \ ,
\end{equation}
with the two-body breakup momentum $q$ and the summed mass $m_{\text{thr}}$ of the final-state particles.
We multiply the amplitude with phase-space and angular-momentum barrier factors and a production factor modelled as in \cite{3pi_rmf}. 
Finally, we multiply the amplitude by a complex-valued coupling that determines the strength and phase of each component.


The free parameters in our model, i.e.\ the nominal masses $m_0$ and widths $\Gamma_0$ of each included resonance, the background parameters $a$ and $b$ and the complex couplings, are estimated by a least-squares fit to the results of the partial-wave decomposition. The fit uses not only the partial-wave intensities, but also interference terms between waves, and their correlations. Note that only a subset of partial waves is modelled in this way, each in a chosen \mX range.
We perform a wide range of systematic studies and estimate the systematic uncertainty of each parameter as the largest observed deviation.

\subsection{Analysis models}

In \KsK, for Pomeron exchange, we expect only resonances with $\JP=(\text{even})^{+}$ \cite{chung_Gparity}.
We hence decompose the data into partial waves with even spins $J \leq 6$, spin projection $M=1$ and $M=2$ for $J=2$. All four waves are included in the resonance-model fit, and are described by six resonances (see \cref{sec:results_aJ}). The $a_2(1320)$ state is parameterized by \cref{eq:BW}, where we include the $\rho\pi$ and $\eta\pi$ decay modes in the dynamic width. For all other states, we set $\Gamma(\mX) = \Gamma_0$ in \cref{eq:BW}.\footnote{In addition, to stabilize the resonance-model fit, we fix the shape of the non-resonant background in the $\JPM=2^{+}1$ wave to a fit to the wave's high-mass tail, where resonant contribution should be negligible.}

We decompose the \oPP data into partial waves with $J\leq8$ and $M \leq2$. The decay chain is parameterized using $\rho(770)$, $b_1(1235)$, $\rho(1450)$, and $\rho_3(1690)$ as isobar resonances using all their known decay channels. A wave-selection procedure based on a regularization technique \cite{flo_wss_proc} is used to determine the set of waves that enters the decomposition. The resonance-model fit covers a subset of 24 waves, measuring 11 resonances with $\JP=0^{-}$, $1^{-}$, $1^{+}$, $2^{-}$, $3^{+}$, $4^{-}$, $4^{+}$ and $6^{+}$.

\section{Measurements of $\JP=(\text{even})^{+}$ mesons}\label{sec:results_aJ}

\begin{table}[]
    \setlength{\tabcolsep}{12pt}
    \renewcommand{\arraystretch}{1.25}
    \centering
    \begin{tabular}{ccll}
    \toprule
    \JP &Resonance & $m_0\ [\MeVcc]$ & $\Gamma_0\ [\MeVcc]$ \\
     \midrule
     \multirow{3}{*}{$2^+$} & $a_2(1320)$ & $1316.63 \pm 0.20 {\,}^{+2.23}_{-2.33}$ & $109.5 \pm 0.4 {\,}^{+2.6}_{-2}$ \\
     & $a_2(1700)$ & $1748 \pm 4 {\,}^{+13}_{-86}$ & $534 \pm 9 {\,}^{+26}_{-230}$ \\
     & $a_2''$ & $2124 \pm 5 {\,}^{+37}_{-9}$ & $527\pm 13 {\,}^{+55}_{-250}$ \\
     \midrule
     \multirow{2}{*}{$4^+$} & $a_4(1970)$ & $ 1952.2 \pm 1.8 {\,}^{+3}_{-3.5}$ & $327 \pm 4 {\,}^{+6}_{-6}$ \\
     & $a_4'$ & $2608 \pm 9 {\,}^{+5}_{-38}$ & $609 \pm 22 {\,}^{+35}_{-311}$ \\
     \midrule
     $6^+$ & $a_6(2450)$ & $2430 \pm 9 {\,}^{+21}_{-25}$ & $523 \pm 22 {\,}^{+39}_{-119}$ \\
     \bottomrule
    \end{tabular}
    \caption{Resonance parameters extracted from the \KsK data, with their statistical and systematic uncertainties respectively.}
    \label{tab:res_parameters_KsK}
\end{table}

\subsection{$\JP=2^{+}$ states}


\begin{figure}
    \centering
    \begin{subfigure}{0.32\linewidth}
        \includegraphics[width=\linewidth]{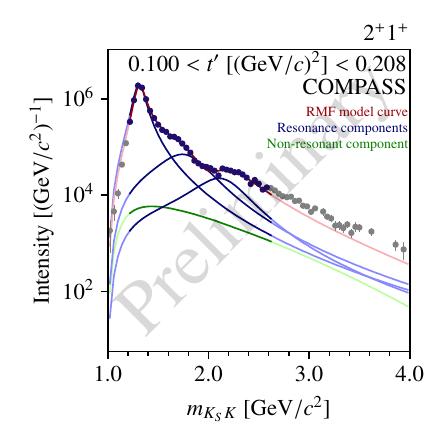}
        \centering
    \end{subfigure}
    \begin{subfigure}{0.32\linewidth}
        \includegraphics[width=\linewidth]{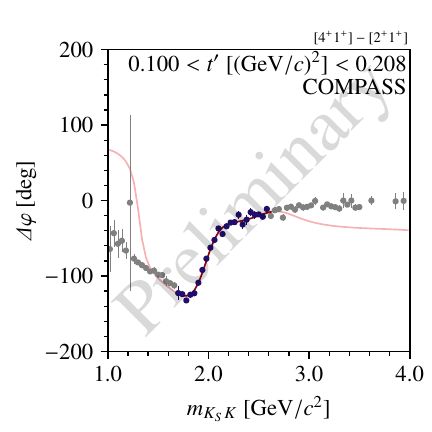}
        \centering
    \end{subfigure}
    \begin{subfigure}{0.32\linewidth}
        \includegraphics[width=\linewidth]{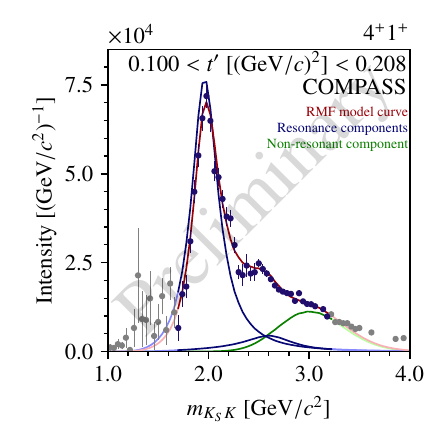}
        \centering
    \end{subfigure}
    \caption{Intensities of the $\JP{}M=2^{+}1$ and $4^{+}1$ waves (left and right), and their relative phase (middle), in \KsK. Greyed-out points are outside the fit range and not taken into account in the resonance-model fit. The red, blue and green curves show the total model and its resonant and non-resonant components, respectively.}
    \label{fig:rmf_2p-4p}
\end{figure}

The intensity distribution of the $\JP{}M=2^{+}1$ wave in the first \tpr bin of the \KsK data is drawn in logarithmic scale in the left plot of \cref{fig:rmf_2p-4p}.
The result of the resonance-model fit is represented by the curves in \cref{fig:rmf_2p-4p}.
Its intensity distribution is clearly dominated by the high, narrow $a_2(1320)$ peak around $1.3\ \GeVcc$. The fitted parameters, listed in \cref{tab:res_parameters_KsK}, are in accordance with previous measurements.

Towards larger masses, the distribution shows a significant shoulder, where we expect the $a_2(1700)$ state to contribute. The shoulder is well described by the interference between the $a_2(1320)$ and the $a_2(1700)$. The latter lies at both a higher mass and a larger width (see \cref{tab:res_parameters_KsK}) than previous measurements, but our estimate is accompanied by large systematic uncertainties.

Around $2.3\ \GeVcc$, we observe another shoulder, which is described in our fit by an additional $a_2''$ state.
No established $a_2$ states are listed in the PDG \cite{PDG} beyond the $a_2(1700)$, but several unconfirmed states are claimed by other experiments. Our resonance-model fit yields clear indications for the existence of such a state.
Our values for the parameters, in \cref{tab:res_parameters_KsK}, are compatible with the state listed as $a_2(2175)$ in \cite{anisovich}. It should be noted that the authors of \cite{anisovich} claim the existence of two more $a_2$ states in this mass region, of which we find no indication.

\subsection{$\JP=4^{+}$ states}

 The intensity of the $4^{+}1$ wave and its phase relative to the $2^{+}1$ wave are shown in the right and middle plots of \cref{fig:rmf_2p-4p}, respectively, for the \KsK final state.
 At $2\ \GeVcc$, we observe a clear peak in the intensity and a distinct phase rise, both well described by a resonance consistent with the well-known $a_4(1970)$ state.
Our precise estimates for its parameters in \KsK (\cref{tab:res_parameters_KsK}) agree with those obtained in the \oPP analysis as well as with earlier measurements. 
 
A large shoulder can be observed at masses above the $a_4(1970)$.
Our fit indicates that, on top of a non-resonant component, an additional $a_4$ state is needed to accurately describe the spectra of the $4^{++}1$ wave. This is not the case in \oPP.
The parameters of this $a_4'$ state are given in \cref{tab:res_parameters_KsK}.


\subsection{$\JP=6^{+}$ states}

\begin{figure}
    \centering
    \begin{subfigure}{0.32\linewidth}
        \includegraphics[width=\linewidth]{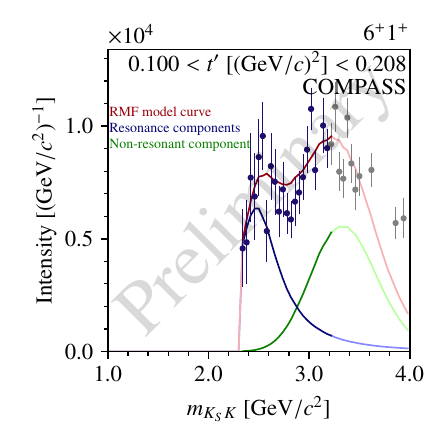}
        \centering
    \end{subfigure}
    \begin{subfigure}{0.32\linewidth}
        \includegraphics[width=\linewidth]{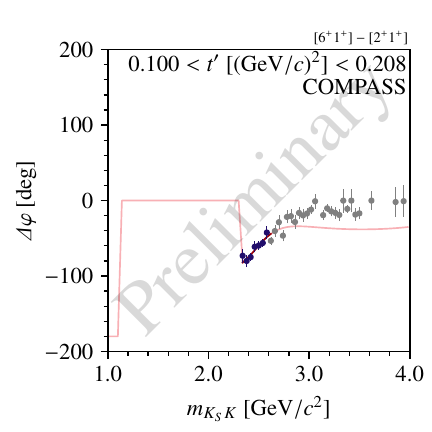}
        \centering
    \end{subfigure}
    \caption{Same as \cref{fig:rmf_2p-4p}, for the $\JPM=6^{+}1$ wave.}
    \label{fig:rmf_6p}
\end{figure}

\Cref{fig:rmf_6p} shows the intensity distribution of the $6^{+}1$ wave in \KsK and its relative phase to the $2^{+}1$ wave.\footnote{To circumvent ambiguities in the partial-wave decomposition \cite{chung_ambiguities}, we only include the wave above $2.32\ \GeVcc$.}
The broad structure centered around $3.4\ \GeVcc$ is described by the non-resonant background.\footnote{We expect significant background in higher $J$ waves from double-Regge exchange \cite{compass_PWA_review}.}
A resonant state, the $a_6(2450)$, is needed to describe the narrower peak-like structure at $2.5\ \GeVcc$.
We also measure the $a_6(2450)$ in \oPP, with consistent parameters.
Only one published measurement \cite{cleland} exists for the $a_6(2450)$ state. It agrees well with our estimates in \cref{tab:res_parameters_KsK}.


\section{\texorpdfstring{Measurements of the spin-exotic $\pi_1(1600)$}{Measurements of the spin-exotic pi\_1(1600)}}\label{sec:results_pi1}

\begin{figure}
    \centering
    \begin{subfigure}{0.32\linewidth}
        \includegraphics[width=\linewidth]{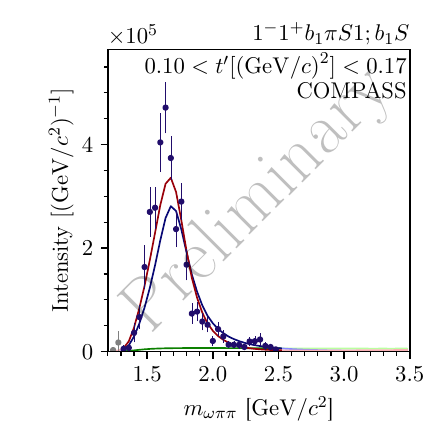}
        \centering
    \end{subfigure}
    \begin{subfigure}{0.32\linewidth}
        \includegraphics[width=\linewidth]{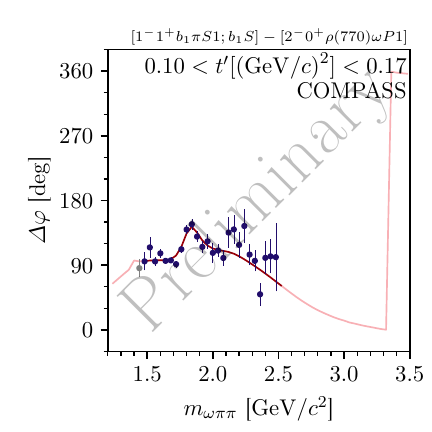}
        \centering
    \end{subfigure}
    \begin{subfigure}{0.32\linewidth}
        \includegraphics[width=\linewidth]{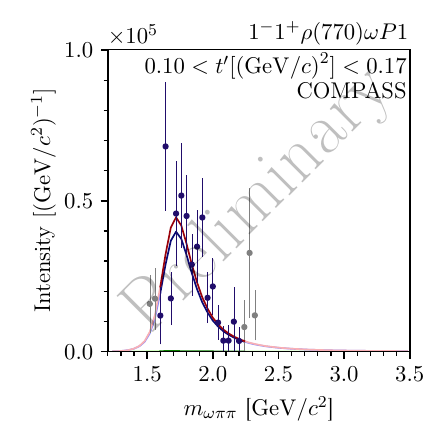}
        \centering
    \end{subfigure}
    \caption{Same as \cref{fig:rmf_2p-4p}, for the $\JPM=1^{-}1$ waves extracted in the $\oPP$ analysis. The left and right plots show the intensities of the decay channels via $b_1(1235)\pi$ and $\rho(770)\omega$, respectively. The middle plot shows the relative phase of the $b_1(1235)\pi$ wave w.r.t.\ a $2^{-}0$ wave decaying via $\rho(770)\omega$.}
    \label{fig:rmf_1mp}
\end{figure}

The analysis of the \oPP final state gives access to spin-exotic $J^{PC}=1^{-+}$ states.
This is of particular interest, as lattice QCD predicts a dominant decay of $\pi_1(1600)$ via $b_1(1235)\pi$~\cite{latticeP1Decays} and a negligible decay via $\rho(770)\omega$, both accessible as intermediate decays in \oPP.
As shown in \cref{fig:rmf_1mp}, we observe a clear resonance-like signal in the decay via $b_1(1235)\pi$ at about $1.6\ \GeVcc$ in both the intensity (left plot) and phase (middle plot).
We also see clear evidence for the $\pi_1(1600)$ decaying to $\rho(770)\omega$ (right plot), which constitutes the first observation of this decay.
Our resonance-model fit describes these signals well using a single Breit-Wigner resonance, parametrized using \cref{eq:BW}.
\footnote{The dynamic width of the $\pi_1(1600)$ in \cref{eq:BW} is parameterized considering the $b_1(1235)\pi$ decay.} 

We measured a nominal mass of $m_0 = 1723 \pm 6 {\,}^{+37}_{-14}\ \MeVcc$ and a width of $\Gamma_0 = 336 \pm  10 {\,}^{+96}_{-33}\ \MeVcc$, which is at tension with the findings of \cite{e852_opp} in the same final state.
This may be explained by the fact that the authors of \cite{e852_opp} also include an excited $\pi_1$ state, which interferes with the $\pi_1(1600)$.
We find no evidence of such a state.
Compared to the COMPASS analysis of $\pi^-\pi^-\pi^+$ \cite{3pi_rmf}, the state is about $120\ \MeVcc$ heavier but still compatible in mass and width.

\section{Summary}

COMPASS has measured the resonance parameters of fifteen $a_J$ and $\pi_J$ states with high statistical precision. Here, we discussed in particular our confirmation of the existence of three high-mass $a_J$ states.
In addition, we have measured the exotic $\pi_1(1600)$ in its decays into $b_1(1235)\pi$ and $\rho(770)\omega$.
It is the first observation of latter decay.

\end{document}